\documentclass[aps,prl,twocolumn,10pt,groupedaddress]{revtex4-1}
\usepackage{amssymb}
\usepackage{graphicx}
\usepackage{amsmath}
\usepackage{hyperref}
\usepackage{subfigure}
\usepackage{float}
\usepackage[utf8]{inputenc}

\begin{document}

\title{Electroweak relaxation of cosmological hierarchy}
\author{Shao-Jiang Wang}
\email{schwang@cosmos.phy.tufts.edu}
\affiliation{Tufts Institute of Cosmology, Department of Physics and Astronomy, Tufts University, 574 Boston Avenue, Medford, Massachusetts 02155, USA}

\begin{abstract}
A simple model for the late-time cosmic acceleration problem is presented in the Starobinsky inflation with a negative bare cosmological constant as well as a nonminimal coupling to the Higgs boson. After electroweak symmetry breaking, the Starobinsky inflaton has been frozen until very recently, becoming a thawing quintessence, and a comparable magnitude to the observed dark energy density can be achieved without fine-tuning. Our proposal essentially reduces the cosmological constant problem into the electroweak hierarchy problem, and its late-time behaviour is also consistent with the recently proposed swampland criteria.
\end{abstract}
\maketitle

\section{Introduction}

Although physics at different energy scales are decoupled from each other according to the renormalization group (RG) flow, the energy scales themselves could still reveal some connections among physics at these scales. It has long been noticed that \cite{Cohen:1998zx} the energy scale of the currently observed dark energy density $\Lambda_\mathrm{DE}^4\sim(10^{-12}\,\mathrm{GeV})^4$ could be expressed as 
\begin{align}\label{eq:UVIR}
\Lambda_\mathrm{DE}^2\sim H_0M_\mathrm{Pl},
\end{align}
where the Planck scale $M_\mathrm{Pl}\sim10^{18}\,\mathrm{GeV}$ is the smallest ultraviolet (UV) length scale and the current Hubble scale $H_0\sim10^{-42}\,\mathrm{GeV}$ is the largest infrared (IR) length scale. This suspicious UV/IR mixing relation has inspired some quests \cite{Hsu:2004ri,Li:2004rb} for the late-time cosmic acceleration problem \cite{Riess:1998cb,Perlmutter:1998np}. The same pattern is also realized for inflation with a form
\begin{align}\label{eq:FLRW}
\Lambda_\mathrm{inf}^2\sim M_\mathrm{Pl}H_\mathrm{inf}
\end{align}
that can be recognized trivially as the Friedmann equation. The face values $\Lambda_\mathrm{inf}\sim10^{16}\,\mathrm{GeV}$ and $H_\mathrm{inf}\sim10^{14}\,\mathrm{GeV}$ could be inferred from the current constraint \cite{Akrami:2018odb} on the tensor-to-scalar ratio $r\lesssim0.01$.

A similar relation was observed recently in \cite{Dimopoulos:2018eam} that
\begin{align}\label{eq:EWinfDE}
v_\mathrm{EW}^2\sim\Lambda_\mathrm{DE}\Lambda_\mathrm{inf},
\end{align}
where the electroweak (EW) hierarchy $v_\mathrm{EW}\sim10^2\,\mathrm{GeV}$ and cosmological hierarchy $\Lambda_\mathrm{DE}^4\sim(10^{-12}\,\mathrm{GeV})^4$ are conspired via an inflationary scale $\Lambda_\mathrm{inf}\sim10^{16}\,\mathrm{GeV}$ without fine-tuning. Therefore, the question of why the observed cosmological constant is extremely small could be transformed into the question of why the measured EW scale is relatively small. The cosmological hierarchy problem can thus be solved as long as a solution to the EW hierarchy problem is known \textit{prior}.

In the spirit of quintessential inflation \cite{Peebles:1998qn}, a concrete example \cite{Dimopoulos:2018eam} to reproduce the relation \eqref{eq:EWinfDE} is constructed in a nonstandard Starobinsky inflation model with a noncanonical kinetic term and a nonstandard Higgs potential,
\begin{align}
&S=\int\mathrm{d}^4x\sqrt{-g}\left\{\frac{M_\mathrm{Pl}^2}{2}R-\frac12\left[1+\left(\frac{h}{v}\right)^2\left(b\frac{M_\mathrm{Pl}}{\phi}\right)^2\right](\partial\phi)^2\right.\nonumber\\
&-\left.\alpha^2M_\mathrm{Pl}^4\left(1-\mathrm{e}^{-\sqrt{\frac23}\frac{\phi}{M_\mathrm{Pl}}}\right)^2-V(h)\left[1+c\left(\mathrm{e}^{-\sqrt{\frac83}\frac{\phi}{M_\mathrm{Pl}}}-1\right)\right]\right\},
\end{align}
where $\alpha=9.97\times10^{-6}$ is fixed from the observation. After Starobinsky inflation and subsequent reheating, the inflaton remnant stays at a minimal slightly shifted from the origin due to a nonzero Higgs potential value $V(0)=\frac{\lambda}{4}v_\mathrm{EW}^4$ in a symmetric phase with $\lambda=0.129$. Once the EW symmetry is broken, the Higgs is relaxed to the EW vacuum and the inflaton is frozen by a dubbed \textit{bait-and-switch} mechanism at a potential energy density
\begin{align}\label{eq:trial}
c^2\frac{V(h=0)^2}{\alpha^2M_\mathrm{Pl}^4}=\frac{c^2\lambda^2v^8}{16\alpha^2M_\mathrm{Pl}^4}=4c^2\times10^{-48}\,\mathrm{GeV}^4
\end{align}
that could match the currently observed dark energy density $\rho_\Lambda\sim2.58\times10^{-47}\,\mathrm{GeV}^4$ for $c\approx2.5$ without fine-tuning. After that, the frozen inflaton starts rolling down a quintessential potential when the Hubble parameter drops down to its current value. Unfortunately, the model in \cite{Dimopoulos:2018eam} has the serious drawback that, at the quantum level, the noncanonical kinetic term $(h/v)^2(\partial\phi/\phi)^2$ is only suppressed by the EW scale, leaving observable signals that would otherwise have been detected in the Higgs decay channels a long time ago. Furthermore, the construction in \eqref{eq:trial} seems highly nontrivial and unnatural.

In this paper, a simple and natural model to reproduce the relation \eqref{eq:EWinfDE} is constructed in the standard Starobinsky inflation model \cite{Starobinsky:1979ty} with a negative bare cosmological constant as well as a nonminimal coupling to the Higgs boson. The general picture of \cite{Dimopoulos:2018eam} is retained \textit{without} the use of any noncanonical kinetic term and nonstandard Higgs potential. The comparable magnitude to the observed dark energy density can be achieved without fine-tuning thanks to the relation \eqref{eq:EWinfDE}. Our proposal is also consistent with the recently proposed swampland criteria \cite{Brennan:2017rbf,Obied:2018sgi,Ooguri:2018wrx} due to the transformed role of the inflaton as a thawing quintessence at late time.

\section{The model}

The action of our proposal in the Jordan frame is 
\begin{align}\label{eq:AJ}
S_J=\int\mathrm{d}^4x\sqrt{-g}&\left[\frac{M^2}{2}\left(1+\frac{R}{8\alpha^2M^2}+\frac{\xi h^2}{M^2}\right)R-\Lambda_b^4\right.\nonumber\\
&\left.-\frac12(\partial h)^2-V(h)-\mathcal{L}_\mathrm{SM+DM}\right],
\end{align}
where $M$ is an unknown energy scale for $R^2$ gravity to be fixed later; $\alpha$ is an inflationary parameter to be fixed by observation; $\xi$ is a nonminimal coupling of the Higgs field $h$ to the Ricci scalar $R$ that eventually will be generated at loop order even if it is absent at tree level \cite{Freedman:1974gs}; $\Lambda_b$ is a bare cosmological constant that turns out to be negative later; and $V(h)$ is the usual Higgs potential of the form
\begin{align}
V(h)&=\begin{cases}
 \frac{\lambda}{4}h^4+\frac{\lambda}{4}v^4, & \hbox{symmetric phase};\\
 \frac{\lambda}{4}(h^2-v^2)^2, & \hbox{broken phase},
\end{cases}
\end{align}
with $\lambda=0.13$. The Lagrangian $\mathcal{L}_\mathrm{SM+DM}$ for the SM along with an unknown dark matter (DM) sector will be left implicitly thereafter. See, e.g., \cite{Wang:2017fuy,Ema:2017rqn,He:2018gyf} for similar actions but in different contexts, and in particular \cite{Akrami:2017cir} for a comprehensive study on $\alpha$-attractor quintessential inflation.

The Starobinsky scalaron $s$ is introduced as an auxiliary field to rewrite \eqref{eq:AJ} as
\begin{align}
S_J=\int\mathrm{d}^4x\sqrt{-g}&\left[\frac{M^2}{2}\left(1+\frac{s}{4\alpha^2M^2}+\frac{\xi h^2}{M^2}\right)R-\frac{s^2}{16\alpha^2}\right.\nonumber\\
&\left.-\Lambda_b^4-\frac12(\partial h)^2-V(h)\right],
\end{align}
so that its equation-of-motion (EOM) $s=R$ could recover the original form \eqref{eq:AJ}. Einstein gravity is recovered at $(s=s_0,h=v)$ if
\begin{align}
M^2=M_\mathrm{Pl}^2-\frac{s_0}{4\alpha^2}-\xi v^2,
\end{align}
and the action in the Jordan frame becomes
\begin{align}
S_J=\int\mathrm{d}^4x\sqrt{-g}&\left[\frac{M_\mathrm{Pl}^2}{2}\Omega(s,h)^2R-\frac12(\partial h)^2\right.\nonumber\\
&\left.-\frac{s^2}{16\alpha^2}-\Lambda_b^4-V(h)\right],
\end{align}
where a conformal factor
\begin{align}
\Omega(s,h)^2=1+\frac{s-s_0}{4\alpha^2M_\mathrm{Pl}^2}+\xi\frac{h^2-v^2}{M_\mathrm{Pl}^2}
\end{align}
is introduced to transform the metric to be $\widetilde{g}_{\mu\nu}=\Omega^2g_{\mu\nu}$ so that the action in the Einstein frame is obtained as of the form
\begin{align}
S_E=&\int\mathrm{d}^4x\sqrt{-\widetilde{g}}\left[\frac{M_\mathrm{Pl}^2}{2}\widetilde{R}-\frac12\left(M_\mathrm{Pl}\sqrt{\frac32}\ln\Omega^2\right)^2\right.\nonumber\\
&\left.-\frac12\frac{(\widetilde{\partial}h)^2}{\Omega^2}-\frac{s^2}{16\alpha^2\Omega^4}-\frac{\Lambda_b^4+V(h)}{\Omega^4}\right].
\end{align}
By introducing the scalar fields $\psi_h=M_\mathrm{Pl}\sqrt{\frac32}\ln\Omega(s,h)^2$, the action in the Einstein frame could be expressed as of the form
\begin{align}\label{eq:AE0}
S_E&=\int\mathrm{d}^4x\sqrt{-\widetilde{g}}\left\{\frac{M_\mathrm{Pl}^2}{2}\widetilde{R}-\frac12(\widetilde{\partial}\psi_h)^2-\frac12\mathrm{e}^{-\sqrt{\frac23}\frac{\psi_h}{M_\mathrm{Pl}}}(\widetilde{\partial}h)^2\right.\nonumber\\
&-\alpha^2M_\mathrm{Pl}^4\left[1-\mathrm{e}^{-\sqrt{\frac23}\frac{\psi_h}{M_\mathrm{Pl}}}\left(1+\xi\frac{h^2-v^2}{M_\mathrm{Pl}^2}-\frac{s_0}{4\alpha^2M_\mathrm{Pl}^2}\right)\right]^2\nonumber\\
&\left.-\mathrm{e}^{-\sqrt{\frac83}\frac{\psi_h}{M_\mathrm{Pl}}}\left[\Lambda_b^4+V(h)\right]\right\}.
\end{align}
As you will see, $\phi\equiv\psi_0$ is the inflaton before EW symmetry breaking and $\varphi\equiv\psi_v$ is the quintessence after EW symmetry breaking. For the sake of simplicity, we will get rid of the tilde symbol and use the following short notations,
\begin{align}
&\omega_h^2=1+\xi\frac{h^2-v^2}{M_\mathrm{Pl}^2},\,\omega_0^2=1-\frac{\xi v^2}{M_\mathrm{Pl}^2},\, \omega_v^2=1;\\
&S=\frac{s}{4\alpha^2M_\mathrm{Pl}^2},\,S_0=\frac{s_0}{4\alpha^2M_\mathrm{Pl}^2},\,\Omega_h^2(S)=S-S_0+\omega_h^2;\\
&U_h=\frac{\Lambda_b^4+V(h)}{\alpha^2M_\mathrm{Pl}^4},\,U_0=\frac{\Lambda_b^4}{\alpha^2M_\mathrm{Pl}^4}+\frac{V(0)}{\alpha^2M_\mathrm{Pl}^4}\equiv U_v+V_0,
\end{align}
to express the action in the Einstein frame as
\begin{align}\label{eq:AE}
S_E=&\int\mathrm{d}^4x\sqrt{-g}\left[\frac{M_\mathrm{Pl}^2}{2}R-\frac12(\partial\psi_h)^2-\frac12\Omega_h^{-2}(\partial h)^2\right.\nonumber\\
&\left.-\alpha^2M_\mathrm{Pl}^4\frac{U_h+S^2}{(S-S_0+\omega_h^2)^2}\right],
\end{align}
where the potential term in the second line will be denoted as $W(S,h)$.

\section{Starobinsky inflation}

To have a successful Starobinsky inflation before EW symmetry breaking, $\omega_0^2-S_0$ in \eqref{eq:AE0} should be positive, and thus $\omega_0^2-S_0=\exp\left(\sqrt{\frac23}c\right)$ for some constant $c$. Furthermore, the bare cosmological constant term in \eqref{eq:AE} should not interfere with the end of inflation roughly at $\phi_\mathrm{end}/M_\mathrm{Pl}=1+c$, namely,
\begin{align}
(\Lambda_b^4+V(0))\mathrm{e}^{-2\sqrt{\frac23}(1+c)}\ll\alpha^2M_\mathrm{Pl}^4\left(1-\mathrm{e}^{-\sqrt{\frac23}}\right)^2,
\end{align}
leading to a constraint
\begin{align}\label{eq:inflation1}
\frac{U_0}{(\omega_0^2-S_0)^2}\ll\left(\mathrm{e}^{\sqrt{\frac23}}-1\right)^2\approx1.6,
\end{align}
that will be checked later. Another constraint comes from the suppression of fluctuations in the Higgs sector to reserve the inflationary prediction of Starobinsky inflation. This requires the effective mass of the kinetically normalized Higgs $\chi$ from $(\mathrm{d}h/\mathrm{d}\chi)^2=\Omega_h^2$ to be larger than the inflationary Hubble scale,
\begin{align}\label{eq:inflation2}
m_\chi^2&=\Omega_0^2W''_h(S,0)=\frac{-4\xi}{M_\mathrm{Pl}^2}\left(\alpha^2M_\mathrm{Pl}^4\frac{S^2}{\Omega_0^4(S)}+\frac{\Lambda_b^4+V(0)}{\Omega_0^4(S)}\right);\nonumber\\
&\approx\frac{-4\xi S^2}{\Omega_0^4(S)}\alpha^2M_\mathrm{Pl}^2\gg H_\mathrm{inf}^2=\frac{S^2}{3\Omega_0^4(S)}\alpha^2M_\mathrm{Pl}^2,
\end{align}
namely, $|\xi|\gg1/12$, which will also be checked later. For now, we will assume that these two constraints are satisfied so that Starobinsky inflation can proceed as usual.

After Starobinsky inflation and subsequent reheating, the inflaton remnant, if it does not decay away totally, stays at a local minimum $(S_\mathrm{EW},h=0)$ determined from the condition
\begin{align}
\frac{W'_\phi(S_\mathrm{EW},0)}{\alpha^2M_\mathrm{Pl}^4}=\sqrt{\frac83}\frac{1}{M_\mathrm{Pl}}\frac{S_\mathrm{EW}(\omega_0^2-S_0)-U_0}{\Omega_0^4(S_\mathrm{EW})}=0,
\end{align}
which gives rise to the field value of the Starobinsky scalaron just before EW symmetry breaking,
\begin{align}
S_\mathrm{EW}=\frac{U_0}{\omega_0^2-S_0}.
\end{align}

\section{EW symmetry breaking}
When EW symmetry is broken, the Higgs is relaxed to its current minimum $h=v$ and the potential energy density is of the form
\begin{align}\label{eq:fixing}
W(S_\mathrm{EW},v)=\alpha^2M_\mathrm{Pl}^4\frac{U_v+S_\mathrm{EW}^2}{(1-S_0+S_\mathrm{EW})^2}.
\end{align}
To retain the success of the picture observed in \cite{Dimopoulos:2018eam}, $\varphi$ should be frozen right after EW symmetry breaking by requiring a light effective mass of $\varphi$,
\begin{align}\label{eq:freezing}
m^2_\varphi&=W''_\varphi(S_\mathrm{EW},v);\nonumber\\
&=\frac43\alpha^2M_\mathrm{Pl}^2\frac{2U_v+(1-S_0)(1-S_0-S_\mathrm{EW})}{(1-S_0+S_\mathrm{EW})^2}.
\end{align}
If the observation from relation \eqref{eq:EWinfDE} indeed reveals the myth of dark energy, all we have to do is to solve the fixing condition \eqref{eq:fixing} and freezing condition \eqref{eq:freezing}, namely,
\begin{align}
a V_0^2&=\frac{U_v+S_\mathrm{EW}^2}{(1-S_0+S_\mathrm{EW})^2};\label{eq:fixed}\\
b V_0^2&=\frac{2U_v+(1-S_0)(1-S_0-S_\mathrm{EW})}{(1-S_0+S_\mathrm{EW})^2}\label{eq:frozen}.
\end{align}
To match the currently observed dark energy density $W(S_\mathrm{EW},v)\sim\Lambda_\mathrm{DE}^4$ and thawing behaviour $m^2_\varphi\sim H_0^2$, one only needs for the order-of-one parameters $a=25/4$ and $b=a/4\Omega_\Lambda$, with $\Omega_\Lambda\approx0.7$ today.

Solving \eqref{eq:fixed} and \eqref{eq:frozen} is a nontrivial task. The only freedom comes from the normalized scalaron value $S_0$, where Einstein gravity is fixed. By choosing $S_0$ away from $1$, one expects following approximated solutions
\begin{align}
\omega_0^2&\approx\frac32S_0-\frac12;\label{eq:approxsol1}\\
U_v&\approx-\frac14(S_0-1)^2.\label{eq:approxsol2}
\end{align}
However, these solutions do not allow for the desirable behaviour at both early time and late time that necessarily requiring $\omega_0^2-S_0>0$ and $1-S_0>0$ from \eqref{eq:AE0}. 

\section{Thawing quintessence}

It turns out as a nice surprise that, when $S_0$ is close to $1^-$, the position of $S_0$ with desirable solutions is independent of the parameters $a$ and $b$. To see this, one could take a concrete example by choosing $S_0=1-V_0$ without lost of generality. The equations \eqref{eq:fixed} and \eqref{eq:frozen} are solved to give
\begin{align}
\omega_0^2&=3-\frac32V_0+(3b-6a)V_0^2+\mathcal{O}(V_0^3);\label{eq:solution1}\\
U_v&=-\frac14V_0^2+\frac34(a+b)V_0^4+\mathcal{O}(V_0^6).\label{eq:solution2}
\end{align}
that are truncated at the order when parameters $a$ and $b$ first appear. The leading order terms of \eqref{eq:solution1} and \eqref{eq:solution2} are indeed independent of the choice of how close $S_0$ is to $1^-$. 

\begin{figure}
\centering
\includegraphics[width=0.45\textwidth]{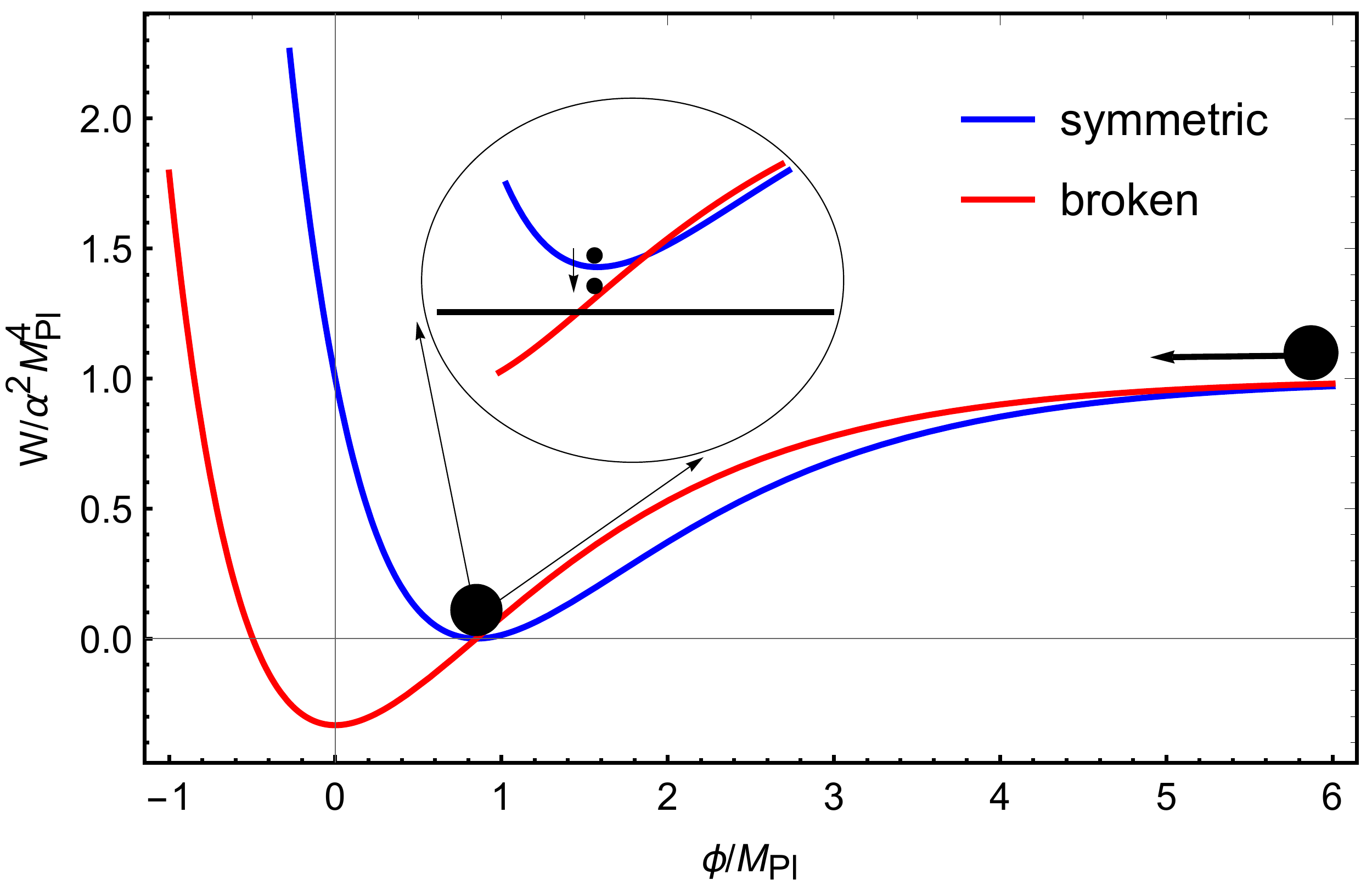}\\
\caption{The demonstration of the physical picture of our model from the truncated solutions \eqref{eq:solution1} and \eqref{eq:solution2} with $S_0=1-V_0$. The potential in \eqref{eq:AE0} along the $\varphi$ direction in the broken phase has been shifted appropriately for illustration.}\label{fig:inflatonDE}
\end{figure}

The potentials in \eqref{eq:AE0} along the $\psi_h$ direction in the symmetric and broken phase are presented in Fig.\ref{fig:inflatonDE} with above truncated solutions from $S_0=1-V_0$, where the EW symmetry breaking occurs in the normalized scalaron value $S_\mathrm{EW}=V_0/2+(6a-3b)V_0^3/4+\mathcal{O}(V_0^5)$, with $\Omega_0^2(S_\mathrm{EW})=2+(3b-6a)V_0^2+\mathcal{O}(V_0^3)$ and $\Omega_v^2(S_\mathrm{EW})=3V_0/2+(6a-3b)V_0^3/4+\mathcal{O}(V_0^5)$, namely, $\phi_\mathrm{EW}=\sqrt{3/2}\ln\Omega_0^2(S_\mathrm{EW})=0.8489M_\mathrm{Pl}$ and $\varphi_\mathrm{EW}=\sqrt{3/2}\ln\Omega_v^2(S_\mathrm{EW})=-151.788M_\mathrm{Pl}$. The broken-phase potential is thus shifted appropriately for clarity. The normalized scalaron value at final AdS minimum in the broken phase is $S_\mathrm{min}=-V_0/4+3(a+b)V_0^3/4+\mathcal{O}(V_0^5)$ with $\Omega_v^2(S_\mathrm{min})=3V_0/4+3(a+b)V_0^3/4+\mathcal{O}(V_0^5)$, namely $\varphi_\mathrm{min}=\sqrt{3/2}\ln\Omega_v^2(S_\mathrm{min})=-152.637M_\mathrm{Pl}$. Note that the rolling of $\varphi$ in the future $\Delta\varphi=\varphi_\mathrm{EW}-\varphi_\mathrm{min}=\phi_\mathrm{EW}$ is a sub-Planckian field excursion.

Using the truncated solutions \eqref{eq:solution1} and \eqref{eq:solution2}, one can check that the original equations \eqref{eq:fixed} and \eqref{eq:frozen} are trivially satisfied at the leading order,
\begin{align}
&\frac{U_v+S_\mathrm{EW}^2}{(1-S_0+S_\mathrm{EW})^2}=a V_0^2+\frac{4a+b}{12}V_0^3+\mathcal{O}(V_0^4);\\
&\frac{\frac{2U_v}{(1-S_0)^2}+1-\frac{S_\mathrm{EW}}{1-S_0}}{\left(1+\frac{S_\mathrm{EW}}{1-S_0}\right)^2}=b V_0^2-\frac{4a+b}{12}V_0^3+\mathcal{O}(V_0^4),
\end{align}
which freezes the inflaton at the right position after EW symmetry breaking with a potential energy density and effective mass
\begin{align}
W(S_\mathrm{EW},v)&=a\frac{V(0)^2}{\alpha^2M_\mathrm{Pl}^4}\approx\Lambda_\mathrm{DE}^4;\label{eq:DE}\\
m_\varphi^2(S_\mathrm{EW},v)&=\frac{4b}{3M_\mathrm{Pl}^2}\frac{V(0)^2}{\alpha^2M_\mathrm{Pl}^4}\approx H_0^2,\label{eq:H0}
\end{align}
desirable for our purpose. The Starobinsky inflaton is thus frozen until the Hubble parameter drops down to its current value and becoming a thawing quintessence today, which also explains the coincidence problem.

One can also check the early-time behaviour from the truncated solutions \eqref{eq:solution1} and \eqref{eq:solution2}. During inflation, the constraint \eqref{eq:inflation1} is explicitly satisfied;
\begin{align}
\frac{U_0}{(\omega_0^2-S_0)^2}=\frac14V_0+\frac{1}{16}V_0^2+\mathcal{O}(V_0^3)\ll1.6.
\end{align}
The other constraint \eqref{eq:inflation2}, or equivalently $|\xi|\gg1/12$, is also explicitly satisfied from \eqref{eq:solution1}, namely,
\begin{align}
-\xi=\frac{M_\mathrm{Pl}^2}{v^2}\left(2-\frac32V_0+\mathcal{O}(V_0^2)\right)\approx10^{32}\gg\frac{1}{12}.
\end{align}
Note that the constraint \cite{Atkins:2012yn} on $|\xi|\lesssim10^{15}$ is not applicable here due to the presence of $R^2$ gravity in addition to the nonminimal coupling. The decay channel of the Higgs to quintessence from the coupling term $\exp(-\sqrt{\frac23}\varphi/M_\mathrm{Pl})(\partial h)^2$ is highly suppressed by the Planck scale, leaving no trace in the collider. The Planckian suppressed effect on various couplings in the SM potential also evades the bounds from the fifth force. The large effective mass of the Higgs during inflation could protect it from the dangerous quantum kick into the unwanted large-field minimum. The Higgs instability problem (see, e.g., \cite{Espinosa:2018mfn,Markkanen:2018pdo} for a brief review) is thus cured as a by-product.

\section{swampland criteria}

The standard single-field slow-roll inflationary paradigm currently faces some tension \cite{Agrawal:2018own} with the original de Sitter conjecture in the swampland criteria \cite{Brennan:2017rbf,Obied:2018sgi} as well as the refined de Sitter conjecture \cite{Ooguri:2018wrx} (see also \cite{Garg:2018reu}) that either one of the following conditions,
\begin{align}
|\nabla V|&\geq\frac{c}{M_\mathrm{Pl}}V;\\
\min(\nabla_i\nabla_j V)&\leq-\frac{c'}{M_\mathrm{Pl}^2}V,
\end{align}
is fulfilled for some universal constants $c, c'>0$ of order 1. Here $V$ is a potential of scalar fields $\phi_i$ in a low energy effective theory of any consistent quantum gravity, and the minimum eigenvalue in the second condition is taken for the Hessian operator $\nabla_i\nabla_j V$ in an orthonormal frame.
See also \cite{Achucarro:2018vey,Garg:2018reu,Brahma:2018hrd,Das:2018hqy,Lin:2018kjm,Motaharfar:2018zyb,Ashoorioon:2018sqb,Das:2018rpg} for possible ways out of swampland, \cite{Lehners:2018vgi,Dias:2018ngv,Denef:2018etk,Colgain:2018wgk,Matsui:2018bsy,Ben-Dayan:2018mhe,Chiang:2018jdg,Heisenberg:2018yae,Kinney:2018nny,Han:2018yrk,Odintsov:2018zai,Reece:2018zvv,Heisenberg:2018rdu,Murayama:2018lie,Marsh:2018kub,Choi:2018rze,Matsui:2018xwa,Hamaguchi:2018vtv,Kawasaki:2018daf} for the implications, and \cite{Danielsson:2018ztv,Andriot:2018wzk,Kehagias:2018uem,Roupec:2018mbn,Andriot:2018ept,Conlon:2018eyr,Dasgupta:2018rtp,Cicoli:2018kdo,Akrami:2018ylq,Bena:2018fqc} for the debates.

Although our action \eqref{eq:AJ} contains a bare cosmological constant, which turns out to be mildly negative deduced from \eqref{eq:solution2},
\begin{align}
\Lambda_b^4\approx-\frac14\frac{V(0)^2}{\alpha^2M_\mathrm{Pl}^4}\sim-\Lambda_\mathrm{DE}^4,
\end{align}
the plateau potential is currently in tension with the swampland criteria, unless turning to, for example,  warm inflation \cite{Das:2018rpg,Motaharfar:2018zyb} or non-Bunch-Davies initial states \cite{Brahma:2018hrd,Ashoorioon:2018sqb}. Nevertheless, the late-time behaviour of our proposal is consistent with the original de Sitter conjecture in the swampland criteria due to the transformed role of the Starobinsky inflaton as a thawing quintessence with
\begin{align}
M_\mathrm{Pl}\frac{|\nabla_{\varphi} W(S_\mathrm{EW},v)|}{W(S_\mathrm{EW},v)}&=\sqrt{\frac83}\frac{(1-S_0)S_\mathrm{EW}-U_v}{S_\mathrm{EW}^2+U_v}\nonumber\\
&\approx\frac{2}{3a}\sqrt{\frac23}V_0^{-2}\gg\mathcal{O}(1),
\end{align}
while
\begin{align}
M_\mathrm{Pl}^2\frac{\min(\nabla_i\nabla_j W)}{W(S_\mathrm{EW},v)}&=M_\mathrm{Pl}^2\frac{\nabla_\varphi\nabla_\varphi W}{W(S_\mathrm{EW},v)}\nonumber\\
&=\frac43\frac{(1-S_0)(1-S_0-S_\mathrm{EW})+2U_v}{S_\mathrm{EW}^2+U_v}\nonumber\\
&\approx\frac{4b}{3a}=\frac{1}{3\Omega_\Lambda}.
\end{align}

The future destiny of our Universe is starting to roll down the quintessential potential, eventually crossing the zero point of the potential and inevitably approaching the final AdS minimum with potential energy density
\begin{align}
W(S_\mathrm{min},v)=\alpha^2M_\mathrm{Pl}^4\frac{U_v+\left(\frac{U_v}{1-S_0}\right)^2}{\left(1-S_0+\frac{U_v}{1-S_0}\right)^2}\approx-\frac13\alpha^2M_\mathrm{Pl}^4
\end{align}
within one Planckian field excursion, $\Delta\varphi=\varphi_\mathrm{EW}-\varphi_\mathrm{min}=\phi_\mathrm{EW}\approx0.85M_\mathrm{Pl}$, which is also consistent with the distance conjecture of the swampland criteria \cite{Brennan:2017rbf,Obied:2018sgi}. 

\section{Conclusion}
To naturally reproduce the conspired relation among the interplay of the EW scale and the inflationary scale with the dark energy scale, we propose a simple model of quintessential Starobinsky inflation to address the late-time cosmic acceleration problem. The model in the Jordan frame is simply defined in $R^2$ gravity with a bare negative cosmological constant term as well as a nonminimal coupling of the Higgs to the Ricci scalar. When transformed into the Einstein frame, the Starobinsky inflation is obtained, and the Higgs instability problem is solved due to a large effective mass. After EW symmetry breaking, the Starobinsky inflaton is frozen at a potential energy density \eqref{eq:DE} comparable to the currently observed dark energy density without fine-tuning. Only until recently when the Hubble parameter drops down to its current value does the inflaton start rolling down a quintessential potential, eventually ending up in an AdS state within one Planckian field excursion. The late time behaviour is consistent with the recently proposed swampland criteria.

\section{Discussion}
There are infinite truncated solutions to Eqs. \eqref{eq:fixed} and \eqref{eq:frozen} with similar leading order terms on the right-hand-side when $S_0$ is close to $1^-$, which might be regarded as a reflection of the string landscape at the effective-field-theory (EFT) level. Any solution with $S_0$ chosen to be away from $1$ is in the regime of string swampland, where our observable Universes cannot be obtained. Even in the regime of the string landscape, a larger value of the EW scale than the currently measured value would either freeze the inflaton at such a large energy density that life cannot have enough time to form or be incapable of freezing the inflaton at all so that our Universe quickly rolls down to the final AdS minimum. Therefore, the anthropic principle for EW hierarchy problem is thus implied. In this respect, although the cosmological constant problem can be naturally solved in light of relation \eqref{eq:EWinfDE} within our model, an explanation for a relatively small EW scale is still needed, for example, supersymmetry \cite{Dimopoulos:1981zb}, extra dimensions \cite{Randall:1999ee,ArkaniHamed:1998rs}, strong dynamics \cite{Weinberg:1975gm,Susskind:1978ms}, cosmological relaxion \cite{Graham:2015cka}, or $N$naturalness \cite{Arkani-Hamed:2016rle}.

\begin{acknowledgments}
We are grateful to Alexander Vilenkin, Zhong-Zhi Xianyu and Masaki Yamada for their useful and stimulating discussions.  We also want to thank Tommi Markkanen for helpful correspondence. This work is supported by the postdoctoral scholarship of Tufts University.
\end{acknowledgments}

\bibliographystyle{utphys}
\bibliography{ref}

\end{document}